\documentclass[english,secnumarabic,amssymb,nobibnotes,aps,prd]{revtex4}
\usepackage[T1]{fontenc}
\usepackage[latin9]{inputenc}
\usepackage{amsmath}
\usepackage{graphicx}
\usepackage{amssymb}

\makeatletter

\usepackage{amsfonts}
\usepackage{epsfig}

\usepackage{psfrag}

\usepackage{colordvi}
\usepackage{color}

\newcommand{\sfrac}[2]{\left[\begin{matrix} #1\\ #2 \end{matrix}\right]}

\makeatother

\begin{document}

\title{Generalized Transformation for Decorated Spin Models}

\author{Onofre Rojas, J. S. Valverde and S. M. de Souza}

\affiliation{Departamento de Ci\^{e}ncias Exatas, Universidade Federal de Lavras,
CP 3037, 37200-000, Lavras - MG, Brazil.}

\begin{abstract}
The paper discusses the transformation of decorated Ising models into an
effective \textit{undecorated} spin models, using the most general
Hamiltonian for interacting Ising models including a long range and high order
interactions. The inverse of a Vandermonde matrix with equidistant nodes $[-s,s]$ is used to obtain an analytical
expression of the transformation. This kind of transformation is very
useful to obtain the partition function of decorated systems. The
method presented by Fisher is also extended, in order to obtain the
correlation functions of the decorated Ising models transforming into
an effective undecorated Ising models. We apply this transformation
to a particular mixed spin-(1/2,1) and (1/2,2) square lattice with
only nearest site interaction. This model could be transformed into
an effective uniform spin-$S$ square lattice with nearest and next-nearest
interaction, furthermore the effective Hamiltonian also include combinations
of three-body and four-body interactions, particularly we considered
spin 1 and 2.
\end{abstract}
\maketitle
\sloppy

\section{Introduction}

Two-leg transformation was first introduced by Fisher\cite{fisher}
in the 1950's decade, after which this transformation was used widely
for one-dimensional decorated models, such as discussed recently for
the tetramer Ising-Heisenberg bond-alternating chain as a model system
for Cu(3-Chloropyridine)$_{2}$(N$_{3}$)$_{2}$ in reference \cite{1d-decor}.
Another decorated model has also been considered recently\cite{1d-decor-e},
one that can be applied even for two-dimensional decorated Ising models\cite{2d-decor}
or higher dimensions.

The Ising models with multi-spin interactions have been extensively
investigated both theoretically and experimentally. In this sense,
we present a review of the start-triangle transformation\cite{fisher}.
This kind of transformation also was discussed in detail by Syozi\cite{Domb}.
This transformation is widely used for two-dimensional models, the
most well known of these being the Ising kagom\'{e} lattice\cite{naya},
the honeycomb lattice\cite{baxter}, and other two-dimensional models
like Ising model on pentagonal lattice\cite{urumov}.

Another illustration of the application of this transformation was
performed for the Ising-Heisenberg diamond chain\cite{1d-decor-e,dmd-IHm}.
This transformation can also be applied to higher dimension decorated
lattice such as, two-dimensional decorated Ising-Heisenberg models\cite{2d-decor}
and two-dimensional doubly decorated Ising-Heisenberg models\cite{2d-ddecor}.

Recently several real systems motivate to investigate in this kinds
of transformation, such as the discovered two-dimensional magnetic
materials $Cu_{9}X_{2}(cpa)_{6}.xH_{2}O$ (cpa=2-carboxypentonic acid;
X=F, Cl, Br) where the $Cu$ spins stands on the triangular kagom\'{e}
lattice\cite{norman} with Heisenberg interaction type. Liquid crystals
networks composed by pentagonal, square and triangular cylinders\cite{bin-chen}.
Other recent investigation about the crystal structure of solvated
{[}Zn(tpt)2/3(SiF,)(H20)2- (MeOH)] {[}tpt = 2,4,6-tris(4-pyridyl)-1,3,5-triazine]
networks with the (10,3)-a topology\cite{robson}.

This paper is organized as follows. In section 2, we start with a
review of star-triangle transformation\cite{fisher} where we comment
their possible extension, afterwards we extend it formally to the general
case of the $m$-leg spin-1/2 system. In section 3 we present the transformation
for the general case of the $m$-leg spin-$S$, using matrix formulation. In
Section 4 we discuss the correlation function for the general case
of the $m$-leg spin-$S$ system also using the matrix formulation. In section
5, this transformation is applied to the mixed spin-(1/2,$S$) of square
Ising lattice with only nearest interaction, spite this model cannot
be mapped into exactly solvable model, we even could discuss their
critical point behavior. Using a double transformation we study the
spin-$S$ square lattice with nearest and next-nearest interaction, furthermore
the effective Hamiltonian also include combinations of three-body
and four-body interactions. Finally in section 6 we present our conclusion.

\section{Transformation of decorated spin-1/2 models}

\subsection{Star-triangle transformation}

\begin{figure}[!ht]

\begin{centering}
\psfrag{s0}{$S_{0}$} \psfrag{s1}{$\sigma_{1}$} \psfrag{s2}{$\sigma_{2}$}
\psfrag{s3}{$\sigma_{3}$} \includegraphics[width=12cm,height=3.5cm]{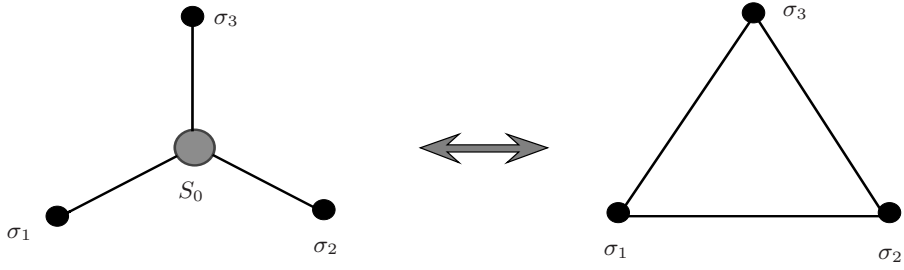} 
\par\end{centering}

\caption[fig2]{Schematic cell representation of three-leg decorated Ising model (left)
and undecorated Ising model (right) or currently known as star-triangle transformation}

\label{fig_2} 
\end{figure}

The star-Ising model with an \textit{arbitrary decorated spin} $S_{0}$, and the spin-1/2 of legs $\sigma_{1}$, $\sigma_{2}$
and $\sigma_{3}$ is presented in the figure 1. Conveniently the Hamiltonian will be defined from now in units of $\beta=1/kT$, being $k$ the Boltzmann constant and $T$ the absolute temperature. Therefore the Hamiltonian is given by 

\begin{align}
{\mathcal{H}}^{(3)}(S_{0},\sigma_{1},\sigma_{2},\sigma_{3})= & J_{0,0,0}^{(3)}S_{0}+J_{1,0,0}^{(3)}S_{0}\sigma_{1}+J_{0,1,0}^{(3)}S_{0}\sigma_{2}+J_{0,0,1}^{(3)}S_{0}\sigma_{3}+J_{1,1,0}^{(3)}S_{0}\sigma_{1}\sigma_{2}+\nonumber \\
& +J_{1,0,1}^{(3)}S_{0}\sigma_{1}\sigma_{3}+J_{0,1,1}^{(3)}S_{0}\sigma_{2}\sigma_{3}+J_{1,1,1}^{(3)}S_{0}\sigma_{1}\sigma_{2}\sigma_{3},\end{align}
 here the parameter $J_{0,0,0}^{(3)}$ represents the magnetic field
acting on spin $S_{0}$, $J_{1,0,0}^{(3)}$, $J_{0,1,0}^{(3)}$ and $J_{0,0,1}^{(3)}$ are the exchange parameter between $S_{0}$ and $\sigma_{1}$, $\sigma_{2}$ and $\sigma_{3}$ respectively. The parameters
of the three-spin interaction are $J_{1,1,0}^{(3)}$, $J_{1,0,1}^{(3)}$
and $J_{0,1,1}^{(3)}$ which corresponds to the three spins products
$S_{0}\sigma_{1}\sigma_{2}$, $S_{0}\sigma_{1}\sigma_{3}$ and $S_{0}\sigma_{2}\sigma_{3}$
respectively. Finally the interaction parameter $J_{1,1,1}^{(3)}$
corresponds to the four-body interaction $S_{0}\sigma_{1}\sigma_{2}\sigma_{3}$.
It is important to notice that the first fourth terms of the Hamiltonian
already was considered early by Fisher\cite{fisher}.

Now the Hamiltonian of the triangle-Ising model represented in Fig.\ref{fig_2} (right) can also be expressed, in
a very general way, as follow 
\begin{align}
\widetilde{\mathcal{H}}^{(3)}(\sigma_{1},\sigma_{2},\sigma_{3})= & \widetilde{J}_{0,0,0}^{(3)}+\widetilde{J}_{1,0,0}^{(3)}\sigma_{1}+\widetilde{J}_{0,1,0}^{(3)}\sigma_{2}+\widetilde{J}_{0,0,1}^{(3)}\sigma_{3}+\widetilde{J}_{1,1,0}^{(3)}\sigma_{1}\sigma_{2}+\nonumber \\
& +\widetilde{J}_{1,0,1}^{(3)}\sigma_{1}\sigma_{3}+\widetilde{J}_{0,1,1}^{(3)}\sigma_{2}\sigma_{3}+\widetilde{J}_{1,1,1}^{(3)}\sigma_{1}\sigma_{2}\sigma_{3},
\end{align}
where $\widetilde{J}_{0,0,0}^{(3)}$ correspond to the effective
parameter corresponding to a constant energy, the coefficients of
$\sigma_{i}$ (with $i=1..3$) represents the effective parameter
of a magnetic field, the coefficients of $\sigma_{i}\sigma_{j}$ (with
$\{i,j\}=1..3$) are the standard bilinear coupling effective parameter,
and the coefficient of $\sigma_{1}\sigma_{2}\sigma_{3}$ represent
the effective three-linear interaction. 

Besides adding a magnetic field parameter to the standard star-triangle
transformation proposed by Fisher\cite{fisher}, we also included
the three-linear parameter, in order to solve the algebraic system
equation consistently. We can verify that there are eight equations
and eight parameters to be obtained.

Carrying out a partial trace over the variable of the decorating system, it is reduced to an effective partition function with spin-1/2 Ising model. Therefore we write down the partition function for a given decorated system 
\begin{align}
\mathcal{Z}(\beta)={\rm e}^{N_{d}\widetilde{J}_{0,0,0}^{(3)}}\widetilde{\mathcal{Z}}(\beta),
\end{align}
 where $N_{d}$ is the number of decorations of the lattice. Considering
a partial summation on decorated particles ($S_{0}$), we have the
following amount, which we will call the associated Boltzmann weight
\begin{align}
W^{(3)}(\sigma_{1},\sigma_{2},\sigma_{3})={\rm tr}_{S_{0}}\Big({\rm e}^{\mathcal{H}^{(3)}(S_{0},\sigma_{1},\sigma_{2},\sigma_{3})}\Big).\label{F3_dcr}\end{align}

On the other hand the Boltzmann factor $\widetilde{W}^{(3)}(\sigma_{1},\sigma_{2},\sigma_{3})$
in the transformed (undecorated) Hamiltonian, becomes 
\begin{align}
\widetilde{W}^{(3)}(\sigma_{1},\sigma_{2},\sigma_{3})= & \exp\big(\widetilde{J}_{0,0,0}^{(3)}+\widetilde{J}_{1,0,0}^{(3)}\sigma_{1}+\widetilde{J}_{0,1,0}^{(3)}\sigma_{2}+\widetilde{J}_{0,0,1}^{(3)}\sigma_{3}+\widetilde{J}_{1,1,0}^{(3)}\sigma_{1}\sigma_{2}+\nonumber \\
& +\widetilde{J}_{1,0,1}^{(3)}\sigma_{1}\sigma_{3}+\widetilde{J}_{0,1,1}^{(3)}\sigma_{2}\sigma_{3}+\widetilde{J}_{1,1,1}^{(3)}\sigma_{1}\sigma_{2}\sigma_{3}\big).\label{F3_udcr}
\end{align}
 The transformation of the parameter of $\widetilde{J}_{n_{1},n_{2},n_{3}}^{(3)}$
into a function of $J_{n_{1},n_{2},n_{3}}^{(3)}$ is obtained, relating
the $W^{(3)}(\sigma_{1},\sigma_{2},\sigma_{3})=\widetilde{W}^{(3)}(\sigma_{1},\sigma_{2},\sigma_{3})$.
Therefore from eqs.\eqref{F3_dcr} and \eqref{F3_udcr}, we get the
solution of the new parameters as a function of the decorated Hamiltonian
parameters, given by \begin{align}
\widetilde{J}_{n_{1},n_{2},n_{3}}^{(3)}=\frac{1}{8}\sum_{\sigma_{1},\sigma_{2},\sigma_{3}=\pm1}\sigma_{1}^{n_{1}}\sigma_{2}^{n_{2}}\sigma_{3}^{n_{3}}\ln\big(W^{(3)}(\sigma_{1},\sigma_{2},\sigma_{3})\big),\label{sol-3leg}\end{align}
 where $n_{i}=0,1$ with $i=1,2,3$.

If we consider the \textit{arbitrary decorated spin} $S_{0}$ as a
simple spin-1/2, this transformation is known as duality of the star-triangle
relation\cite{baxter} and is useful to solve other spin models such
as kagom\'{e} like models.

The solution showed in eq.\eqref{sol-3leg} corresponds to the eight
solutions. A particular case of this solution becomes the solution
obtained by Fisher\cite{fisher}, when we consider $\widetilde{J}_{1,0,0}^{(3)}=\widetilde{J}_{0,1,0}^{(3)}=\widetilde{J}_{0,0,1}^{(3)}=0$
and $\widetilde{J}_{1,1,1}^{(3)}=0$, leaving thus only four free
parameters to be determined instead of eight.

This particular solution is widely applied to a large variety of two
dimensional lattice. In a recent paper, Lackov\'{a} \textit{et al.}\cite{mixed-triang}
discussed an exact results of a mixed spin-1/2 and spin-1 transverse
Ising model with two- and four-spin interactions and crystal field
on the honeycomb lattice.

\subsection{m-leg star-polygon transformation}

The transformation of the decorated model presented by Fisher\cite{fisher}
could also be extended to the \textit{cross-square} transformation
as considered recently by Savvidy\cite{savvidy}, to study the systems
with exponentially degenerated vacuum state. A particular case of this
transformation was discussed in reference \cite{savvidy}, with the
bilinear and the four-linear terms as the non-null parameters, thus
maintaining only three free parameters in the Hamiltonian, which is
necessary to study the gonihedric model\cite{gonihedric}.

Motivated by the previous result, we can extended the formulation
to a general case, where we consider $m$ particles interacting with
a central \textit{spin} $S_{0}$. This kind of model can be used to
study two or three dimensional lattices with high order coordination
number. The Hamiltonian of the decorated model with non-linear interaction and central
spin $S_{0}$ can be expressed as %
\begin{figure}[!ht]
\begin{centering}
\psfrag{s0}{$S_{0}$} \psfrag{s1}{$\sigma_{1}$} \psfrag{s2}{$\sigma_{2}$}
\psfrag{s3}{$\sigma_{3}$} \psfrag{s4}{$\sigma_{4}$} \psfrag{si}{$\sigma_{i}$}
\psfrag{sm}{$\sigma_{m}$} \includegraphics[width=12cm,height=4.5cm]{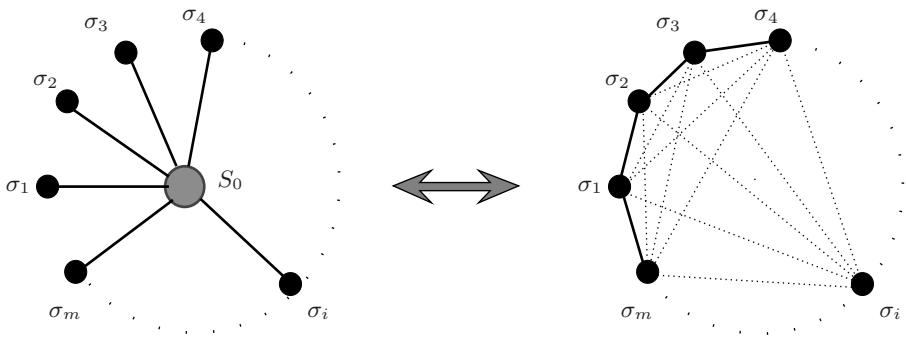} 
\par\end{centering}
\caption[fig4]{Schematic representation of m-leg decorated Ising model (left) and
m-side polygon Ising model (right)}
\label{fig_4} 
\end{figure}

\begin{align}
\mathcal{H}^{(m)}(S_{0},\{\sigma_{i}\}_{m})=\sum_{\{n_{i}\}_{m}=0,1}J_{\{n_{i}\}_{m}}^{(m)}S_{0}\prod_{i=1}^{m}\sigma_{i}^{n_{i}},\label{m-legH}
\end{align}
where $J_{\{n_{i}\}_{m}}^{(m)}$ is the parameter interaction of
$\prod_{i=1}^{m}\sigma_{i}^{n_{i}}$ with sub-indexes $\{\sigma_{i}\}_{m}$.
We denote the set $\{\sigma_{i}\}_{m}=\{\sigma_{1},\dots,\sigma_{m}\}$,
similarly by $\{n_{i}\}_{m}=\{n_{1},\dots,n_{m}\}$, whereas the super
indexes are related to the number of legs or coordination number of
decorated spin.

On the other hand, the undecorated Hamiltonian into which the decorated
Hamiltonian \eqref{m-legH} will be transformed, is given by 
\begin{align}
\widetilde{\mathcal{H}}^{(m)}(\{\sigma_{i}\}_{m})=\sum_{\{n_{i}\}_{m}=0,1}\widetilde{J}_{\{n_{i}\}_{m}}^{(m)}\prod_{i=1}^{m}\sigma_{i}^{n_{i}}.\label{mlegH-half}
\end{align}
The transformation of the parameters $\widetilde{J}_{\{n_{i}\}_{m}}^{(m)}$
as a function of $J_{\{n_{i}\}_{m}}^{(m)}$ are obtained in a similar
way as in the previous case. Writing the associated Boltzmann weight,
as a trace over the spin-$S_{0}$ of the decorated system, 
\begin{align}
W^{(m)}(\{\sigma_{i}\}_{m})={\rm tr}_{S_{0}}\Big({\rm e}^{\mathcal{H}^{(3)}(S_{0},\{\sigma_{i}\}_{m})}\Big).\label{Fm_dcr-g}
\end{align}

On the other hand the Boltzmann weight function $W^{(m)}(\{\sigma_{i}\}_{m})$, for the transformed Hamiltonian can be expressed as 
\begin{align}
\widetilde{W}^{(m)}(\{\sigma_{i}\}_{m})=\prod_{\{\sigma_{i}\}_{m}=\pm1}\exp\big(\widetilde{J}_{\{n_{i}\}_{m}}^{(m)}\prod_{i=1}^{m}\sigma_{i}^{n_{i}}\big).\label{Fm_udcr}
\end{align}
Assuming that $W^{(m)}(\{\sigma_{i}\}_{m})=\widetilde{W}^{(m)}(\{\sigma_{i}\}_{m})$ and eq.\eqref{Fm_dcr-g} are known, we substitute them in eq.\eqref{Fm_udcr}, getting the solution of the unknown parameters which reads as 
\begin{align}
\widetilde{J}_{\{n_{i}\}_{m}}^{(m)}=\frac{1}{2^{m}}\sum_{\{\sigma_{i}\}_{m}=\pm1}\big(\prod_{i=1}^{m}\sigma_{i}^{n_{i}}\big)\ln\big(W^{(m)}(\{\sigma_{i}\}_{m})\big).\label{sol-mleg}
\end{align}
Rewritten the eq.\eqref{sol-mleg} a little different, we have 
\begin{align}
\widetilde{J}_{\{n_{i}\}_{m}}^{(m)}=\sum_{\{\sigma_{i}\}_{m}=\pm1}g_{\{n_{i}\}_{m}}(\{\sigma_{i}\}_{m}\big)\ln\big(W^{(m)}(\{\sigma_{i}\}_{m})\big),\label{g_nm}
\end{align}
where the function $g_{\{n_{i}\}_{m}}$ depends only of spins $\sigma_{i}$,
\begin{align}
g_{\{n_{i}\}_{m}}(\{\sigma_{i}\}_{m})=\frac{1}{2^{m}}\prod_{i=1}^{m}\sigma_{i}^{n_{i}},\end{align}
 with $n_{i}=0,1$ and $i=1,2,\dots,m$. Eq.\eqref{g_nm} corresponds
to $2^{m}$ solutions of the unknown parameters $\widetilde{J}_{\{n_{i}\}_{m}}^{(m)}$.

Let us consider a special case as an example, without losing its general
properties, suppose we consider the central spin $S_{0}=\pm1/2$,
then the eq.\eqref{Fm_dcr-g} is reduced to \begin{align}
W^{(m)}(\{\sigma_{i}\}_{m})=2\cosh\big(\sum_{\{n_{i}\}_{m}=0,1}J_{\{n_{i}\}_{m}}^{(m)}\prod_{i=1}^{m}\sigma_{i}^{n_{i}}\big).\label{Fm_dcr}\end{align}

\section{Transformation for spin-$s$ models}

One interesting extension considered here is the higher spins of decorated models, then for this purpose we first consider the two-leg spin-1 model, where we introduce a matrix formalism to simplify our notation. After that we will discuss the fully general spin-$s$ m-leg star-polygon transformation.

\subsection{Two-leg transformation for spin-1 model}

In this section we discuss the spin-1 case. Recently similar situation
was discussed by Fireman \textit{et al.}\cite{fireman} where they
consider the Ising model with bi-linear, bi-quadratic, single ion
anisotropy and Zeeman interaction, which was mapped into an effective
Blume Emery Griffiths (BEG) model\cite{beg}. Extending to the spin-1
model, it is possible to write the Hamiltonian according to the definition
in eq.\eqref{m-legH}, \begin{align}
\mathcal{H}^{(2)}(S_{0},s_{1},s_{2})=\sum_{n_{1},n_{2}=0}^{2}J_{n_{1},n_{2}}^{(2)}S_{0}s_{1}^{n_{1}}s_{2}^{n_{2}},\label{ham-3m}\end{align}
 where $J_{n_{1},n_{2}}^{(2)}$ corresponds to interaction parameter,
here $s_{1}$ and $s_{2}$ represents the spin-1, whereas $S_{0}$
could be any mechanical system. The effective Hamiltonian of the undecorated
system also can be expressed in analogy with eq.\eqref{mlegH-half},
which would give us, \begin{align}
\widetilde{\mathcal{H}}^{(2)}(s_{1},s_{2})=\sum_{n_{1},n_{2}=0}^{2}\widetilde{J}_{n_{1},n_{2}}^{(2)}s_{1}^{n_{1}}s_{2}^{n_{2}},\label{ham-3m-udr}\end{align}
 with $\widetilde{J}_{n_{1},n_{2}}^{(2)}$ being the interaction parameters
of the transformed system. Considering a partial summation of the
partition function where involves the summation (trace) on spin $S_{0}$, we obtain as before the so called associated Boltzmann weight,
given by,
\begin{align}
W^{(2)}(s_{1},s_{2})={\rm tr}_{S_{0}}\Big({\rm e}^{\mathcal{H}^{(2)}(S_{0},s_{1},s_{2})}\Big).\label{F_dcr1}
\end{align}
To solve the unknown parameter we introduce the matrix formalism which is an appropriate representation to express these equations in terms of the known parameters. Using the matrix notation we write the eq.\eqref{ham-3m}, simply as follow
\begin{align}
\mathsf{H}^{(2)}(S_{0})=S_{0}{\bf V}^{(1,1)}\mathsf{J}^{(2)},\label{lin-eqMat}\end{align}
 where ${\bf V}^{(1,1)}$ and $\mathsf{J}^{(2)}$ are defined respectively
by \begin{align}
{\bf V}^{(1,1)}=\begin{pmatrix}1 & -1 & 1 & -1 & 1 & -1 & 1 & -1 & 1\\
1 & 0 & 0 & -1 & 0 & 0 & 1 & 0 & 0\\
1 & 1 & 1 & -1 & -1 & -1 & 1 & 1 & 1\\
1 & -1 & 1 & 0 & 0 & 0 & 0 & 0 & 0\\
1 & 0 & 0 & 0 & 0 & 0 & 0 & 0 & 0\\
1 & 1 & 1 & 0 & 0 & 0 & 0 & 0 & 0\\
1 & -1 & 1 & 1 & -1 & 1 & 1 & -1 & 1\\
1 & 0 & 0 & 1 & 0 & 0 & 1 & 0 & 0\\
1 & 1 & 1 & 1 & 1 & 1 & 1 & 1 & 1\end{pmatrix}\quad\text{and}\quad\mathsf{J}^{(2)}=\begin{pmatrix}J_{0,0}^{(2)}\\
J_{0,1}^{(2)}\\
J_{0,2}^{(2)}\\
J_{1,0}^{(2)}\\
J_{1,1}^{(2)}\\
J_{1,2}^{(2)}\\
J_{2,0}^{(2)}\\
J_{2,1}^{(2)}\\
J_{2,2}^{(2)}\end{pmatrix}.\end{align}

To transform the decorated Hamiltonian in eq.\eqref{ham-3m} into an undecorated Hamiltonian, as given in eq.\eqref{ham-3m-udr} we find the parameters of effective Hamiltonian as a function of the original decorated Hamiltonian. Therefore, we define the associated Boltzmann weight function $W^{(2)}(s_{1},s_{2})$ as 
\begin{align}
W^{(2)}(s_{1},s_{2})={\rm e}^{\widetilde{\mathcal{H}}^{(2)}(s_{1},s_{2})}.\label{F_udcr1}
\end{align}

To make a complete matrix representation, we define the following function $\mathcal{R}^{(2)}(s_{1},s_{2})=\ln(W^{(2)}(s_{1},s_{2}))$. The function $\mathcal{R}^{(2)}(s_{1},s_{2})$ can be expressed as a vector $\mathsf{R}^{(2)}$.
Using the matrix notation and considering $\widetilde{W}^{(2)}(s_{1},s_{2})=W^{(2)}(s_{1},s_{2})$,
we have the following equation 
\begin{align}
\mathsf{R}^{(2)}=\mathbf{V}^{(1,1)}\widetilde{\mathsf{J}}^{(2)}.\label{eq: Rsp-1}
\end{align}
The solution of the equation above could be obtained by taking the
inverse of the matrix ${\mathbf{V}}^{(1,1)}$, denoted from now on
by $\widetilde{\mathbf{V}}^{(1,1)}=(\mathbf{V}^{(1,1)})^{-1}$, The
solution is then expressed by 
\begin{align}
\widetilde{\mathsf{J}}^{(2)}=\widetilde{\mathbf{V}}^{(1,1)}\mathsf{R}^{(2)}.\label{tata}
\end{align}
But the $\mathbf{V}^{(1,1)}$ can be written also as $\mathbf{V}^{(1,1)}=\mathsf{V}^{(1)}\otimes\mathsf{V}^{(1)}$ and its inverse becomes $\widetilde{\mathbf{V}}^{(1,1)}=\widetilde{\mathsf{V}}^{(1)}\otimes\widetilde{\mathsf{V}}^{(1)}$.
In the present case the matrix $\mathsf{V}^{(1)}$ and $\widetilde{\mathsf{V}}^{(1)}$
could be written explicitly as \begin{equation}
\mathsf{V}^{(1)}=\begin{pmatrix}1 & -1 & 1\\
1 & 0 & 0\\
1 & 1 & 1\end{pmatrix}\quad\text{and}\quad\widetilde{\mathsf{V}}^{(1)}=\begin{pmatrix}0 & 1 & 0\\
-\tfrac{1}{2} & 0 & \tfrac{1}{2}\\
\tfrac{1}{2} & -1 & \tfrac{1}{2}\end{pmatrix}.\end{equation}
 Certainly it is simpler to evaluate the inverse of the reduced matrix
$\mathsf{V}^{(1)}$ instead of the large matrix $\mathbf{V}^{(1,1)}$.

As an example we can consider the non-uniform two-leg Ising spin transformation with central spin $S_{0}=1$, without losing any generality. For this particular case the eq. \eqref{F_dcr1} will be expressed as
\begin{align}
W^{(2)}(s_{1},s_{2})= & 1+2\cosh\big(\sum_{n_{1},n_{2}=0}^{2}J_{n_{1},n_{2}}^{(2)}s_{1}^{n_{1}}s_{2}^{n_{2}}\big).\label{Fg_dcr1-x}
\end{align}

We remark that this transformation is carried out for every unitary cell and can be applied to the one-dimensional \cite{1d-decor-e}, two-dimensional \cite{streckta1} and other high dimensional spin lattices.

\subsection{Two-leg transformation for spin-$s$ model}

For a higher spin the Hamiltonian can be written in a general way
using the previous matrix notation \begin{align}
\mathsf{H}^{(2)}(S_{0})=S_{0}\mathbf{V}^{(s,s)}\mathsf{J}^{(2)}=S_{0}\mathsf{V}^{(s)}\otimes\mathsf{V}^{(s)}\mathsf{J}^{(2)},\label{Hamt-trs-S}\end{align}
 where the dimension of the matrix $\mathbf{V}^{(s,s)}$ is $(2s+1)^{2}\times(2s+1)^{2}$,
meanwhile, the dimension of the matrix $\mathsf{V}^{(s)}$ is $(2s+1)\times(2s+1)$.
We can see that the matrix $\mathsf{V}^{(s)}$ is a Vandermonde matrix
with \textit{equidistant nodes} $[-s,s]$, and the elements of the
nodes are $x_{j}$ which corresponds only to the magnetic moments
of the spin-$s$. The elements can be appropriately expressed as $x_{j}=-s-1+j$,
with $j=1,2,\dots,2s+1$. The explicit representation of the Vandermonde
matrix is given by \begin{align}
\mathsf{V}^{(s)}=\begin{pmatrix}1 & x_{1} & x_{1}^{2} & x_{1}^{3} & \dots & x_{1}^{2s}\\
1 & x_{2} & x_{2}^{2} & x_{2}^{3} & \dots & x_{2}^{2s}\\
\vdots & \vdots & \vdots & \vdots & \ddots & \vdots\\
1 & x_{2s} & x_{2s}^{2} & x_{2s}^{3} & \dots & x_{2s}^{2s}\\
1 & x_{2s+1} & x_{2s+1}^{2} & x_{2s+1}^{3} & \dots & x_{2s+1}^{2s}\end{pmatrix}.\label{vandermonde}\end{align}

Using he matrix notation, the function $\mathsf{R}^{(2)}(s_{1},s_{2})$ is defined in a similar way as was defined in eq.\eqref{eq: Rsp-1}, therefore we have a vector function given by

\begin{align}
\mathsf{R}^{(2)}=\ln(\mathsf{W}^{(2)})=\ln\Big({\rm tr}_{S_{0}}\big({\rm e}^{{\sf H}^{(2)}(S_{0})}\big)\Big).\label{def-RS}
\end{align}

The effective Hamiltonian of the undecorated system also can be written in analogy to eq.\eqref{ham-3m-udr}. Thus we have 
\begin{align}
\widetilde{\mathsf{H}}^{(2)}=\mathbf{V}^{(s,s)}\widetilde{\mathsf{J}}^{(2)}=\mathsf{V}^{(s)}\otimes\mathsf{V}^{(s)}\widetilde{\mathsf{J}}^{(2)}.\label{H2til}
\end{align}

The Hamiltonian considered in eq.\eqref{H2til}, can contain at most
$(2s+1)^{2}$ parameters. This also means that the dimension of the
vectors $\mathsf{J}^{(2)}$ and $\mathsf{R}^{(2)}$ is $(2s+1)^{2}$.
Taking the inverse of the matrix $\mathbf{V}^{(s,s)}$ we are able
to express the new parameters as a function of the known parameters
defined by eq.\eqref{Hamt-trs-S}, the new parameters then are the
elements of 
\begin{align}
\widetilde{\mathsf{J}}^{(2)}=\widetilde{\mathbf{V}}^{(s,s)}\mathsf{R}^{(2)}=\widetilde{\mathsf{V}}^{(s)}\otimes\widetilde{\mathsf{V}}^{(s)}\mathsf{R}^{(2)}.
\end{align}

The inverse of the matrix $\mathsf{V}^{(s)}$ could be solved using
the recursive equation presented recently by Eisinberg \textit{et
al.}\cite{eisinberg}, where is discussed an generic algorithm to
obtain the elements of the inverse of the Vandermonde matrix $\widetilde{\mathsf{V}}^{(s)}$.

Therefore following that algorithm\cite{eisinberg}, we find that
the elements of the matrix $\widetilde{\mathsf{V}}^{(s)}$ could be
written explicitly as 
\begin{align}
\widetilde{\mathsf{V}}_{i,j}^{(s)}=\frac{(-1)^{i+j}}{(2s+1-j)!(j-1)!}\sum_{k=1}^{2s+1}(-s-1)^{k-i}\binom{k}{i}\left|\sfrac{2s+2}{k+1}\right|{\rm F}_{i+1}^{1,i-k}\big(1-\tfrac{j}{s+1}\big),\label{invs-elem}
\end{align}
where $\sfrac{.}{.}$ represents the first kind of Stirling number,
whereas ${\rm F}_{i+1}^{1,i-k}$ represents the hyper-geometric
function as defined in \cite{abramowitz}.

It could be interesting to normalize the spin-s. In these cases, it
is possible to rewrite the elements of the Vandermonde matrix $\mathsf{V}^{(s)}$
with equidistant nodes $[-1,1]$, whose elements are given by ${s}_{j}=(-s-1+j)/s$.
The solution of this case has already be found by Eisinberg \textit{et
al.}\cite{eisinberg}. Here we present the same solution but using
the hyper-geometric function explicitly, as follows \begin{align}
\widetilde{\mathsf{V}}_{i,j}^{(s)}=\frac{(-1)^{i+j}s^{i-1}}{(2s+1-j)!(j-1)!}\sum_{k=1}^{2s+1}(-s-1)^{k-i}\binom{k}{i}\left|\sfrac{2s+2}{k+1}\right|{\rm F}_{i+1}^{1,i-k}\big(1-\tfrac{s(s+1-j)}{s+1}\big),\end{align}
Certainly this solution could be useful when the spin-s is large.

\subsection{The m-leg star-polygon transformation for arbitrary spin}

This section presents the general extension for the $m$-leg star-polygon
transformation of decorated Ising spin model. Based on the previous
results, the considered  Hamiltonian will be written as follows 
\begin{align}
{\sf H}^{(m)}(S_{0})=\bigotimes_{i=1}^{m}{\sf V}^{(s_{i})}S_{0}{\sf J}^{(m)},
\end{align}
where $s_{i}$ corresponds to an arbitrary spin for each leg. Assuming that we have $m$ legs of the star or $m$ edge of the polygon with $S_{0}$ being the central spin, then the dimension of the matrix ${V}^{(s_{i})}$ is $(2s_{i}+1)\times(2s_{i}+1)$. The Hamiltonian that we consider can contain at most $\prod_{i=1}^{m}(2s_{i}+1)$ parameters, which means that the dimension of the vector ${\sf J}^{(m)}$ and ${\sf R}^{(m)}$ is $\prod_{i=1}^{m}(2s_{i}+1)$. The elements of the vector ${\sf J}^{(m)}$ are the parameters of the original system.

We define in analogy to the eq.\eqref{def-RS}, a vector function
for the following amount 
\begin{align}
{\sf R}^{(m)}=\ln\Big({\rm tr}_{S_{0}}\big({\rm e}^{{\sf H}^{(m)}(S_{0})}\big)\Big).\label{def-RSg}
\end{align}
The effective Hamiltonian in an undecorated system also can be expressed in analogy to \eqref{H2til}, thus we have 
\begin{align}
\widetilde{\sf H}^{(m)}=\bigotimes_{i=1}^{m}{\sf V}^{(s_{i})}{\widetilde{\sf J}}^{(m)},
\end{align}
where the ${\widetilde{\sf J}}^{(m)}$ is a vector whose elements
are the parameters of the transformed system. Therefore, the new parameters
could be written using the vector representation as 
\begin{align}
{\widetilde{\sf J}}^{(m)}=\bigotimes_{i=1}^{m}{\widetilde{\mathsf{V}}}^{(s_{i})}{\sf R}^{(m)}\label{deltaHF},
\end{align}
we remark that the matrix ${\widetilde{\mathsf{V}}}^{(s_{i})}$ is the inverse of the matrix ${\mathsf{V}}^{(s_{i})}$. The corresponding Boltzmann weigth could be written as follow

\begin{equation}
W(\{s_{m}\})=\mathrm{e}^{\widetilde{H}(\{s_{i}\})}.\label{peso_w}\end{equation}

The explicit form of matrix ${\mathsf{V}}^{(s)}$ and its respective inverse is presented in appendix A, up to spin-3. For an arbitrary spin-s, the matrix ${\mathsf{V}}^{(s)}$ is given by eq.\eqref{vandermonde} and the elements of the matrix inverse are given by the eq.\eqref{invs-elem}.

\section{The correlation function of decorated spin models}

The correlation function in the transformed Hamiltonian involving
the central spin $S_{0}$ can be obtained in a very similar way such
as initially proposed by Fisher\cite{fisher}. Here we present a
very general extension to obtain the correlation function, using the
correlation function of effective systems (undecorated systems).

The correlation function involving spins such as $S_{0},s_{k_{1}},\dots,s_{k_{r}}$,
with $s_{k_{j}}$ any arbitrary spin of system, can be written as
\begin{align}
\langle S_{0}s_{k_{1}}\dots s_{k_{r}}\rangle=\frac{1}{\mathcal{Z}}{\rm tr}_{S_{0},\{s_{i}\}}\Big(S_{0}s_{k_{1}}\dots s_{k_{r}}{\rm e}^{{\mathcal{H}}^{(m)}(S_{0},s_{1},\dots,s_{m})}\Big),\label{gen-corr}\end{align}
where the trace is performed over all $S_{0}$ and $\{s_{i}\}=\{s_{1},s_{2},\dots,s_{m}\}$,
whereas ${\mathcal{Z}}$ is the partition function for the m-leg Ising
model. For simplicity, we consider, as before the Hamiltonian in units
of $\beta=-1/kT$. The partial summation over $S_{0}$ can be expressed
as follows 
\begin{align}
{\mathcal{C}}(s_{1},\dots,s_{m})=\sum_{S_{0}}S_{0}{\rm e}^{{\mathcal{H}}^{(m)}(S_{0},s_{1},\dots,s_{m})}.\label{corr1}
\end{align}

On the other hand we want to write the correlation function \eqref{gen-corr} as a function of all spins of the $m$-leg lattice; for that purpose, we can try to represent the ${\mathcal{C}}(s_{1},\dots,s_{m})$ as a linear combination of $s_{1},\dots,s_{m}$, which reads as 
\begin{align}
{\mathcal{C}}(s_{1},\dots,s_{m})=\sum_{\{n_{i}\}_{m}}\alpha(n_{1},\dots,n_{m})\prod_{i=1}^{m}s_{i}^{n_{i}},\label{corr2}
\end{align}
where $\alpha(n_{1},\dots,n_{m})$ are coefficients to be determined.
Matrix notation is usually very convenient, as was shown in the previous
section. Then we write the eq.\eqref{corr2} using matrix notation,
which reduces it simply to the following expression \begin{align}
{\sf C}^{(m)}=\bigotimes_{i=1}^{m}{\sf V}^{(s_{i})}{\pmb\alpha}^{(m)},\label{corr2v}\end{align}
 where the dimension of vectors ${\sf C}^{(m)}$ and ${\pmb\alpha}^{(m)}$
is $\prod_{i=1}^{m}(2s_{i}+1)$. The elements of the vector ${\sf C}^{(m)}$
are given by eq.\eqref{corr1}, whereas the elements of vector ${\pmb\alpha}^{(m)}$
are the unknown coefficients $\alpha(n_{1},\dots,n_{m})$ to be determined
in a similar way as ${\widetilde{\sf J}}^{(m)}$ was determined by
eq.\eqref{deltaHF}. Therefore the solution of the unknown elements
of the vector ${\pmb\alpha}^{(m)}$ is given by \begin{align}
{\pmb\alpha}^{(m)}=\bigotimes_{i=1}^{m}{\widetilde{\mathsf{V}}}^{(s_{i})}{\sf C}^{(m)}.\end{align}

To obtain the explicit form of the elements of ${\pmb\alpha}^{(m)}$,
the inverse of the same Vandermonde matrix ${\mathsf{V}}^{(s)}$ must
be evaluated for each coordination number. Thus, the correlation function
can be written as a linear combination of the correlation function
of the transformed Hamiltonian, and could be written explicitly as
\begin{align}
\langle S_{0}s_{k_{1}}\dots s_{k_{r}}\rangle=\sum_{\{n_{i}\}_{m}}\alpha(n_{1},\dots,n_{m})\langle s_{k_{1}}\dots s_{k_{r}}\prod_{i=1}^{m}s_{i}^{n_{i}}\rangle.\end{align}

Finally, we are able to write the correlation function including the
decorated spin, which can always be expressed just as the correlation
function of the effective undecorated Ising system.

\section{Critical points of 2D Ising spin-s model with up to quartic interaction}

The two dimensional lattice is one of the most interesting subject
in statistical physics, both experimentally\cite{beg,Bernasconi}
and theoretically. Several approximation methods, such as the mean-field-theory\cite{beg,Capel-Mukamel},
the Bethe approximation\cite{Chakraborty}, the correlated effective
field theory\cite{Kaneyoshi}, the renormalization group\cite{Krinsky-Berker},
the series expansion methods\cite{Soul}, the Monte Carlo methods\cite{Jain}
and the cluster variation methods are used to investigate this interesting
lattice. However, an exact solution has been obtained only in a very
limited cases, mainly in the honeycomb lattices\cite{Horiguchi-Wu,Kolesik}.
Some exact results with restricted parameters has been investigated
by Mi and Yang\cite{MiYang} using a non-one-to-one transformation\cite{Kolesik}.

\begin{figure}[!ht]
\begin{centering}
\psfrag{s0}{$S_{0}$} \psfrag{s1}{$s_{1}$} \psfrag{K}{$K$}
\psfrag{D}{$D$} \psfrag{La}{${\cal L}_{a}$} \psfrag{Lb}{${\cal L}$}
\psfrag{Lc}{${\cal L}_{c}$} \psfrag{s}{$\sigma$} \psfrag{m}{$S$}
\includegraphics[width=8cm,height=7.5cm]{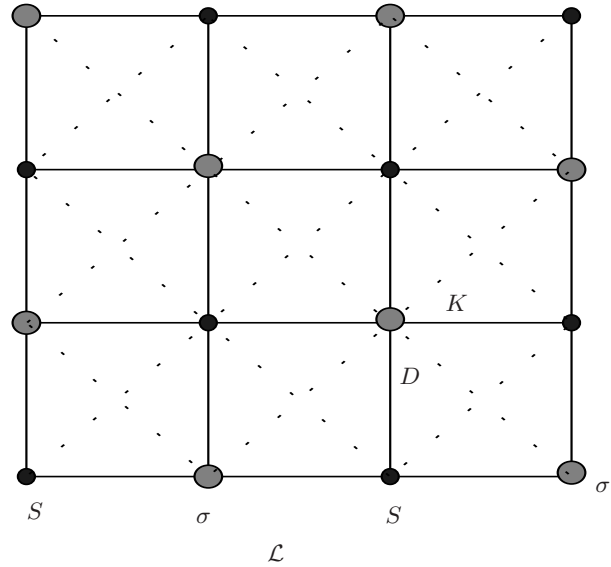} 
\par\end{centering}
\caption[fig5]{Schematic representation of square-type lattice (${\mathcal{L}}$)
with spin-1/2, mixed spin-(1/2,S) on square lattice.}
\label{fig_5} 
\end{figure}

As an example we apply the transformation method presented in the previous section to a lattice with mixed spin-1/2 and spin-$S$. The schematic transformation is displayed in fig.\ref{fig_5}

\subsection{The mixed Ising spin-(1/2,S) lattice}

It is possible to transform a mixed spin lattice into an effective spin-1/2 lattice, such as presented in the literature \cite{fan-wu}. If the spin-$S$ site is considered as a decoration of the lattice ${\cal L}$, then the Hamiltonian is given by 
\begin{align}
{\mathcal{H}}_{1/2,S}=\sum_{<i,j>}KS_{i}\sigma_{j}+\sum_{i}DS_{i}^{2}\label{ising-mixed1}
\end{align}
where $K$ being the first neighbor interaction and $D$ the single
ion-anisotropy parameter. The case when $D=0$ has already been discussed
by Tang\cite{tang}. With $\sigma_{i}$ we represent the spin-1/2
particle, while with $S_{i}$ we represent the spin-S particle.

We have the following Hamiltonian for the effective spin-1/2 Ising model, which is given by

\begin{align}
{\mathcal{H}}_{1/2}=J_{0}+\sum_{(i,j)}J_{2}\sigma_{i}\sigma_{j}+\sum_{\text{all square}}J_{4}\sigma_{i_{1}}\sigma_{i_{2}}\sigma_{i_{3}}\sigma_{i_{4}},\label{ising-half}
\end{align}
where $J_{0}$ means the constant energy, $J_{2}$ corresponds to the interaction parameter with its summation $(i,j)$ running
over all the pairs of spins interactions for each unitary cell of the lattice and $J_{4}$ represents the interaction parameter among all spins for a given unitary cell of squared lattice considered. 

The Boltzmann weight of the eight-vertex model \cite{baxter}, could be obtain using the equation (\ref{peso_w}) as follow 
\begin{align}
w_{1}= & W(\tfrac{1}{2},\tfrac{1}{2},\tfrac{1}{2},\tfrac{1}{2}),\quad w_{2}=W(\tfrac{1}{2},\tfrac{-1}{2},\tfrac{1}{2},\tfrac{-1}{2}), & \quad w_{3}=W(\tfrac{1}{2},\tfrac{-1}{2},\tfrac{-1}{2},\tfrac{1}{2}), & \quad w_{4}=W(\tfrac{1}{2},\tfrac{1}{2},\tfrac{-1}{2},\tfrac{-1}{2}),\nonumber \\
w_{5}= & W(\tfrac{1}{2},\tfrac{-1}{2},\tfrac{1}{2},\tfrac{1}{2}),\quad w_{6}=W(\tfrac{-1}{2},\tfrac{1}{2},\tfrac{1}{2},\tfrac{1}{2}), & \quad w_{7}=W(\tfrac{1}{2},\tfrac{1}{2},\tfrac{-1}{2},\tfrac{1}{2}), & \quad w_{8}=W(\tfrac{1}{2},\tfrac{1}{2},\tfrac{1}{2},\tfrac{-1}{2}).
\end{align}

Due to the symmetry of the lattice, the Boltzmann weights of the mixed spin-(1/2,$S$) lattice are related by 
\begin{align}
w_{2}= & w_{3}=w_{4},\\
w_{5}= & w_{6}=w_{7}=w_{8},
\end{align}

where 
\begin{eqnarray}
w_{1}=\left\{ \begin{array}{lr}
1+2\sum_{n=1}^{S}\cosh(2nK){\rm e}^{n^{2}D}; & S=\text{integral}\\
2\sum_{n=1}^{S+\tfrac{1}{2}}\cosh(nK){\rm e}^{n^{2}D}; & S=\text{half-odd-integral}\end{array}\right.\label{w1s}
\end{eqnarray}
\begin{eqnarray}
w_{2}=\left\{ \begin{array}{lr}
1+2\sum_{n=1}^{S}{\rm e}^{n^{2}D}; & S=\text{integral}\\
2\sum_{n=1}^{S+\tfrac{1}{2}}{\rm e}^{n^{2}D}; & S=\text{half-odd-integral}\end{array}\right.\label{w2s}
\end{eqnarray}
and 
\begin{eqnarray}
w_{5}=\left\{ \begin{array}{lr}
1+2\sum_{n=1}^{S}\cosh(nK){\rm e}^{n^{2}D}; & S=\text{integral}\\
2\sum_{n=1}^{S+\tfrac{1}{2}}\cosh(\tfrac{n}{2}K){\rm e}^{n^{2}D}; & S=\text{half-odd-integral}\end{array}\right.\label{w3s}
\end{eqnarray}

From eqs. \eqref{w1s}, \eqref{w2s} and \eqref{w3s} we can verify the following relation for the Boltzmann factors, $w_{1}>w_{5}>w_{2}$, for any arbitrary values of $K$ and $D$. The transformation of the mixed Hamiltonian \eqref{ising-mixed1} into the effective spin-1/2 Hamiltonian \eqref{ising-half}, relate their parameters, using the Boltzmann weight function, which reads as 
\begin{align}
J_{0}= & \frac{1}{8}\ln\big(w_{2}^{3}w_{5}^{4}w_{1}\big),\quad J_{0}>0,\\
J_{2}= & \frac{1}{8}\ln\Big(\frac{w_{1}}{w_{2}}\Big),\quad J_{2}>0,\\
J_{4}= & \frac{1}{8}\ln\Big(\frac{w_{1}w_{2}^{3}}{w_{5}^{4}}\Big),\quad J_{4}\in{\mathfrak{R}}.
\end{align}

The eight-vertex model has been solved approximately by Fan and Wu \cite{fan-wu}
when $\Delta/w_{m}^{2}\ll1$ with $w_{m}=\text{max}(w_{1},w_{2},w_{3},w_{4})$,
and with 
\begin{align}
\Delta=\omega_{1}\omega_{2}+\omega_{3}\omega_{4}-\omega_{5}\omega_{6}-\omega_{7}\omega_{8}.\label{condition-s}
\end{align}
Unfortunately the free-fermion condition ($\Delta=0$), does not
satisfy our transformed lattice, unless when $D\rightarrow\pm\infty$,
or $T\rightarrow\infty$. But even so, we can discuss their critical
points behavior, using the critical condition 
\begin{align}
w_{1}=\bar{w}_{2}+w_{3}+w_{4},\quad\text{where}\quad\bar{w}_{2}=w_{2}-\Delta/w_{1}.\label{eq:crit}
\end{align}

We have calculated the critical coupling $D_{c}$ as a function of
the parameter $K_{c}$, the curve is displayed in fig.4, for spin-(1/2,1)
and spin-(1/2,2). Using the equation \eqref{eq:crit} in all regions of our plots,
we verified the amount $|\Delta|/w_{1}^{2}<1$, guarantying the convergence
of the approximation. A similar situation was considered by Tang \cite{tang}.

\begin{figure}[!ht]
\begin{centering}
\psfrag{K}{$K_{c}$} \psfrag{D}{$D_{c}$} \psfrag{Spin-1}[][][0.6]{spin-$(1/2,1)$} \psfrag{Spin-2}[][][0.6]{spin-$(1/2,2)$} \includegraphics[width=8cm,height=10cm,angle=-90]{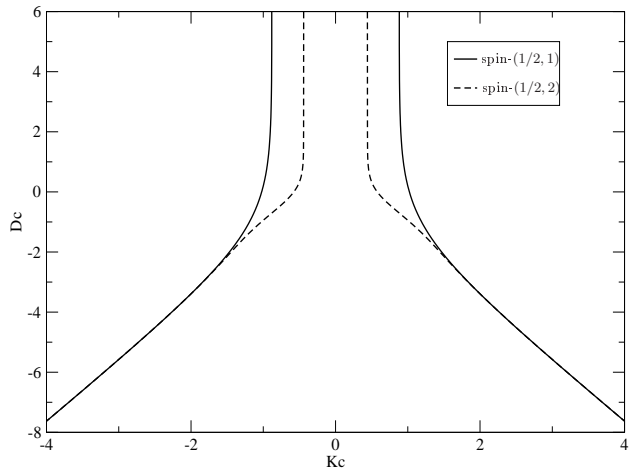} 
\par\end{centering}
\caption[fig5]{We display the critical regions for $D_{c}$ as a function of $K_{c}$ in the case of the free-fermion aproximation. The solid line corresponds to spin-1, while with the dashed line we represent spin-2 lattice.}
\label{fig_6} 
\end{figure}

We can verify that the limiting case of the critical function: when
$K_{c}\rightarrow\pm\infty$ we have a linear behavior of the parameter
$D_{c}=-2|K_{c}|-\ln(2)/2$, whereas when $D_{c}\rightarrow\infty$,
the parameter $K_{c}\rightarrow\pm\ln(1+\sqrt{2})/S$.

Another possibility of the exact solvable condition could be the symmetric vertex model, but unfortunately this model also does not satisfy the symmetric vertex model condition because we have always $w_{1}>w_{2}$ instead $w_{1}=w_{2}$.

We point out that the matrix method calculation of the parameters becomes relevant for higer spin values, the method used in this case simplify effectively very tedious calulations. For example, for spin-1 the calculation of the interaction parameters, without using the matrix method, even could be obtained using the transformation of reference \cite{fisher}. However for spin-2 case it turns on a very cumbersome calculation, we can obtain easily these parameters using the matrix method and algebraic software.

\newpage

\section{Conclusions}

This paper presents an extension of the transformation of decorated spin models presented by Fisher\cite{fisher} into an effective Ising-type models, including high order and long range interaction terms for an arbitrary number of particles and with an arbitrary spin-$s$. We have shown that the general transformation of decorated spin models is reduced to the calculation of the inverse of the especial case of the Vandermonde matrix with equidistant nodes $[-s,s]$. The matrix formulation of decorated spin models transformation could be manipulated more easily when is studied an involving decorated systems. This matrix formulation is the main achievement of this work.

The correlation function of decorated spin models could also be transformed into an effective Ising-type model, using the matrix formulation considered above. We verify that, to obtain the correlation function, it is necessary to perform the inverse of the same Vandermonde matrix with equidistant nodes.

As an application, we present a square-type Ising lattice with mixed spin-1/2 and spin-$S$, following we transform this model into an effective Ising model with spin-1/2. Similar models can involve more complicated calculations in the case of higer spin values where the matrix method proposed in our manuscript turns relevant. In this sense we conclude that the matrix procedure for obtaining the unknown interaction parameters simplify effectively very tedious calculations. Finally, as illustration we give in the appendix explicitly the Vandermonde matrix up to spin-3.

\vskip1cm
\textbf{Acknowledgments}\\
J. S. Valverde thanks FAPEMIG for full financial support, O. Rojas and S.M. de Souza thanks CNPq and FAPEMIG for partial financial support.

\appendix

\section{Some values of the matrix $\mathsf{V}^{(s)}$ and $\widetilde{\mathsf{V}}^{(s)}$}

In this appendix we present the matrix $\mathsf{V}^{(s)}$ and its
inverse $\widetilde{\mathsf{V}}^{(s)}$, for spins up to spin-3. For
spin-3/2 \begin{equation}
\mathsf{V}^{(\frac{3}{2})}=\left(\begin{array}{cccc}
1 & -\frac{3}{2} & \frac{9}{4} & -\frac{27}{8}\\
1 & -\frac{1}{2} & \frac{1}{4} & -\frac{1}{8}\\
1 & \frac{1}{2} & \frac{1}{4} & \frac{1}{8}\\
1 & \frac{3}{2} & \frac{9}{4} & \frac{27}{8}\end{array}\right)\quad\widetilde{\mathsf{V}}^{(\frac{3}{2})}=\left(\begin{array}{cccc}
-\frac{1}{16} & \frac{9}{16} & \frac{9}{16} & -\frac{1}{16}\\
\frac{1}{24} & -\frac{9}{8} & \frac{9}{8} & -\frac{1}{24}\\
\frac{1}{4} & -\frac{1}{4} & -\frac{1}{4} & \frac{1}{4}\\
-\frac{1}{6} & \frac{1}{2} & -\frac{1}{2} & \frac{1}{6}\end{array}\right)\end{equation}
 For spin-2 \begin{equation}
\mathsf{V}^{(2)}=\left(\begin{array}{ccccc}
1 & -2 & 4 & -8 & 16\\
1 & -1 & 1 & -1 & 1\\
1 & 0 & 0 & 0 & 0\\
1 & 1 & 1 & 1 & 1\\
1 & 2 & 4 & 8 & 16\end{array}\right)\quad\widetilde{\mathsf{V}}^{(2)}=\left(\begin{array}{ccccc}
0 & 0 & 1 & 0 & 0\\
\frac{1}{12} & -\frac{2}{3} & 0 & \frac{2}{3} & -\frac{1}{12}\\
-\frac{1}{24} & \frac{2}{3} & -\frac{5}{4} & \frac{2}{3} & -\frac{1}{24}\\
-\frac{1}{12} & \frac{1}{6} & 0 & -\frac{1}{6} & \frac{1}{12}\\
\frac{1}{24} & -\frac{1}{6} & \frac{1}{4} & -\frac{1}{6} & \frac{1}{24}\end{array}\right)\end{equation}
 For spin-5/2 \begin{equation}
\mathsf{V}^{(\frac{5}{2})}=\left(\begin{array}{cccccc}
1 & -\frac{5}{2} & \frac{25}{4} & -\frac{125}{8} & \frac{625}{16} & -\frac{3125}{32}\\
1 & -\frac{3}{2} & \frac{9}{4} & -\frac{27}{8} & \frac{81}{16} & -\frac{243}{32}\\
1 & -\frac{1}{2} & \frac{1}{4} & -\frac{1}{8} & \frac{1}{16} & -\frac{1}{32}\\
1 & \frac{1}{2} & \frac{1}{4} & \frac{1}{8} & \frac{1}{16} & \frac{1}{32}\\
1 & \frac{3}{2} & \frac{9}{4} & \frac{27}{8} & \frac{81}{16} & \frac{243}{32}\\
1 & \frac{5}{2} & \frac{25}{4} & \frac{125}{8} & \frac{625}{16} & \frac{3125}{32}\end{array}\right)\quad\widetilde{\mathsf{V}}^{(\frac{5}{2})}=\left(\begin{array}{cccccc}
\frac{3}{256} & -\frac{25}{256} & \frac{75}{128} & \frac{75}{128} & -\frac{25}{256} & \frac{3}{256}\\
-\frac{3}{640} & \frac{25}{384} & -\frac{75}{64} & \frac{75}{64} & -\frac{25}{384} & \frac{3}{640}\\
-\frac{5}{96} & \frac{13}{32} & -\frac{17}{48} & -\frac{17}{48} & \frac{13}{32} & -\frac{5}{96}\\
\frac{1}{48} & -\frac{13}{48} & \frac{17}{24} & -\frac{17}{24} & \frac{13}{48} & -\frac{1}{48}\\
\frac{1}{48} & -\frac{1}{16} & \frac{1}{24} & \frac{1}{24} & -\frac{1}{16} & \frac{1}{48}\\
-\frac{1}{120} & \frac{1}{24} & -\frac{1}{12} & \frac{1}{12} & -\frac{1}{24} & \frac{1}{120}\end{array}\right)\end{equation}

For spin-3 \begin{equation}
\mathsf{V}^{(3)}=\left(\begin{array}{ccccccc}
1 & -3 & 9 & -27 & 81 & -243 & 729\\
1 & -2 & 4 & -8 & 16 & -32 & 64\\
1 & -1 & 1 & -1 & 1 & -1 & 1\\
1 & 0 & 0 & 0 & 0 & 0 & 0\\
1 & 1 & 1 & 1 & 1 & 1 & 1\\
1 & 2 & 4 & 8 & 16 & 32 & 64\\
1 & 3 & 9 & 27 & 81 & 243 & 729\end{array}\right)\quad\widetilde{\mathsf{V}}^{(3)}=\left(\begin{array}{ccccccc}
0 & 0 & 0 & 1 & 0 & 0 & 0\\
-\frac{1}{60} & \frac{3}{20} & -\frac{3}{4} & 0 & \frac{3}{4} & -\frac{3}{20} & \frac{1}{60}\\
\frac{1}{180} & -\frac{3}{40} & \frac{3}{4} & -\frac{49}{36} & \frac{3}{4} & -\frac{3}{40} & \frac{1}{180}\\
\frac{1}{48} & -\frac{1}{6} & \frac{13}{48} & 0 & -\frac{13}{48} & \frac{1}{6} & -\frac{1}{48}\\
-\frac{1}{144} & \frac{1}{12} & -\frac{13}{48} & \frac{7}{18} & -\frac{13}{48} & \frac{1}{12} & -\frac{1}{144}\\
-\frac{1}{240} & \frac{1}{60} & -\frac{1}{48} & 0 & \frac{1}{48} & -\frac{1}{60} & \frac{1}{240}\\
\frac{1}{720} & -\frac{1}{120} & \frac{1}{48} & -\frac{1}{36} & \frac{1}{48} & -\frac{1}{120} & \frac{1}{720}\end{array}\right)\end{equation}

\end{document}